\documentclass[prd, amsfonts, twocolumn, nofootinbib, showpacs]{revtex4}
\usepackage{graphicx, epsfig}
\usepackage{color}
\usepackage{amsmath}
\usepackage{amssymb}
\newcommand{\be}{\begin{equation}}
\newcommand{\ee}{\end{equation}}
\newcommand{\bea}{\begin{eqnarray}}
\newcommand{\eea}{\end{eqnarray}}

\newcommand{\gapp}{\mathrel{\raise.3ex\hbox{$>$}\mkern-14mu
\lower0.6ex\hbox{$\sim$}}}
\newcommand{\lapp}{\mathrel{\raise.3ex\hbox{$<$}\mkern-14mu
\lower0.6ex\hbox{$\sim$}}}
\def\bbox{{\,\lower0.9pt\vbox{\hrule \hbox{\vrule height 0.2 cm
\hskip 0.2 cm \vrule  height 0.2 cm}\hrule}\,}}

\begin{document}
\title{Pre-Hawking radiation may allow for reconstruction of the mass distribution of the collapsing object}
\author{De-Chang Dai$^1$, Dejan Stojkovic$^2$}
\affiliation{$^1$ Institute of Natural Sciences, Shanghai Key Lab for Particle Physics and Cosmology, \\
and Center for Astrophysics and Astronomy, Department of Physics and Astronomy,\\
Shanghai Jiao Tong University, Shanghai 200240, China}
\affiliation{$^2$
HEPCOS, Department of Physics, SUNY, University at Buffalo, Buffalo, NY 14260-1500}


\begin{abstract}
\widetext
Hawking radiation explicitly depends only on the black hole's total mass, charge and angular momentum. It is therefore generally believed that one cannot reconstruct the information about the initial mass distribution of an object that made the black hole. However, instead of looking at radiation from a static black hole, we can study the whole time-dependent process of the gravitational collapse, and pre-Hawking radiation which is excited because of the time-dependent metric.
We compare radiation emitted by a single collapsing shell with that emitted by two concentric shells of the equivalent total mass. We calculate the gravitational trajectory and the momentum energy tensor. We show that the flux of energy emitted during the collapse by a single shell is significantly different from the flux emitted by two concentric shells of the equivalent total mass. When the static black hole is formed, the fluxes become indistinguishable. This implies that an observer studying the flux of particles from a collapsing object could in principle reconstruct information not only about the total mass of the collapsing object, but also about the mass distribution.

\end{abstract}


\maketitle

\section{Introduction}

The process of black hole formation and subsequent evaporation is associated with some fundamental problems.
Since a stationary black hole solution is characterized by three conserved quantities - mass, charge and angular momentum, and all additional information about the initial state of matter that formed the black hole is lost during the collapse, it may happen that this process implies the fundamental (rather than practical) loss of information \cite{Hawking:1976ra}. Very recently, a new direction was proposed in \cite{Hawking:2016msc} toward the resolution of the information loss paradox. Namely, the authors of   \cite{Hawking:2016msc} argue that the black hole vacuum is not unique, but is actually highly (or perhaps infinitely) degenerate. In addition to the standard conserved quantities, there exist an infinite number of additional quantities  which characterize the black hole solution. These quantities arise from the group of super-translations which acts both near the horizon and at null infinity, and can presumably  map the information imprinted in the near horizon region to the distant observer's region. An important technical detail is that these infinitely degenerate vacua are labeled the same energy but differ in angular momentum. As argued by the authors, this additional angular momentum information might play some role in resolving the information loss paradox.

A question that immediately arises in this context is what happens to information that does not depend on the angular momentum. Generally speaking this includes the global charges (e.g. lepton number, baryon number, flavor) and mass/energy distributions. It was argued in \cite{Stojkovic:2005zq} that violation of global quantum number conservation is not associated with information loss since these numbers are violated even in the absence of the black holes, for example via processes mediated by lepton-quark interactions or the electroweak instantons, and we do not say that these processes imply information loss. However, the question of the mass or energy distribution is more serious. The final black hole state will remember only the total mass of the collapsing object, but not the distribution of masses. For example, one can prepare two different initial states which have the same mass. One state may be a single collapsing spherically symmetric shell of certain mass, and the second state may be made of two concentric shells whose total  mass is equal to the mass of the single shell in the first case. The final states in both cases are the same - a black hole with the same total mass. Since Hawking radiation depends only on the value of the total mass (assuming that the shells carry no charge nor angular momentum), then the information about the mass distribution in the initial state will be lost. It is also not clear at all how would the new proposal in \cite{Hawking:2016msc} help in this case since nothing depends on angular momentum.

This seems to imply that information about the initial state might be released during the collapse, since once the collapse is over there is no much one can do.
It is well known that during the collapse an object radiates away its higher multipoles and other irregularities in the so-called balding phase before a perfect spherically symmetric horizon is formed. However, the situation in our example with two shells is spherically symmetric, so there are no higher multipoles that could be radiated away in the form of gravitational  radiation.
Instead, we could focus on radiation of the fields in the background of the collapsing object which is excited due to the time dependence in the metric during the collapse.
In \cite{Vachaspati:2006ki,Vachaspati:2007hr}, it was shown that gravitational collapse is followed by the so-called pre-Hawking radiation from the very beginning of the collapse, simply because the metric is time dependent. This radiation becomes thermal Hawking radiation only in $t \rightarrow \infty$ limit when the event horizon is formed. Since the collapsing object has only finite amount of mass, an asymptotic observer would never witness the formation of the horizon at $t \rightarrow \infty$. For him, the collapsing object will slowly get converted into not-quite-thermal radiation before it reaches its own Schwarzschild radius. It was demonstrated in \cite{Saini:2015dea} that the evolution is completely unitary in such a setup. It was also argued that the collapsing process may be used to reconstruct information \cite{Lochan:2015oba}.

In this paper, we also concentrate on the pre-Hawking radiation, but we are using the standard analysis of tracing the field modes in the time-dependent gravitational background as defined in \cite{Davies,1984qfcs.book.....B}. We explicitly construct an example in which the initial states are different, i.e. a single shell vs. two concentric shells with the same total mass, while the final state is the same, i.e. a black hole of the same mass. We show that the flux of energy emitted in these two cases is notably different, though in the limit of $t \rightarrow \infty$ the fluxes become identical. Thus, an observer studying the flux of particles from a collapsing object could in principle reconstruct information not only about the total mass of the collapsing object, but also about the mass distribution.

\section{The geometry of the collapsing shells}

\begin{figure}
   \centering
\includegraphics[width=5cm]{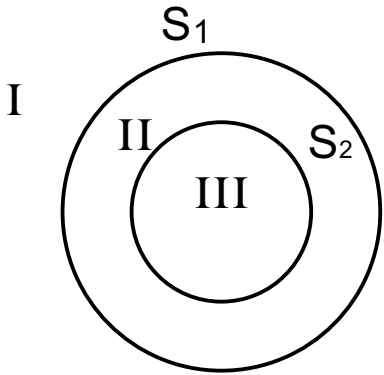}
\caption{$S_1$ and $S_2$ are two freely falling concentric spherical shells. These two shells separate the space into three regions -- I, II, and III. The equations of motion of the shells can be found by matching the geometry inside and outside the shells.}
\label{shell}
\end{figure}

 In this section we consider geometry of two freely falling massive spherically symmetric shells. The shells are concentric, and we label the outer one as $S_1$ and the inner one as $S_2$, as shown in Fig.~\ref{shell}. The rest masses of the shells are $\mu_1$  and $\mu_2$, while their total gravitational masses are $M$ and $m$.  Note that the rest masses and total gravitational masses are not equal here since the shells are moving. This configuration separates the space in three regions labeled by I, II and III. The time dependent radii of the shells $S_1$ and $S_2$ are $R_1$ and $R_2$ respectively. The geometry in the region I  ($r>R_1$) is
\begin{eqnarray}
&&ds^2=-\left(1-\frac{2M}{r}\right)dt_I^2+\left(1-\frac{2M}{r}\right)^{-1} dr^2+r^2d\Omega\\
&&d\Omega=d\theta^2+\sin^2\theta d\phi^2  .
\end{eqnarray}
The geometry in the region II ($R_1>r>R_2$) is
\begin{eqnarray}
ds^2=-\left(1-\frac{2m}{r}\right)dt_{II}^2+\left(1-\frac{2m}{r}\right)^{-1} dr^2+r^2d\Omega
\end{eqnarray}
The geometry in the region III ($r <R_2$) is
\begin{eqnarray}
ds^2=-dt_{III}^2+dr^2+r^2d\Omega
\end{eqnarray}
The time parameters in these three regions are different (otherwise one could not smoothly match the metric at the boundaries of the regions).
The equation of motion of the shells can be found by matching the geometry inside and outside the shells \cite{Lightman}.
These equations are given in terms of the conserved quantities $\mu_1$  and $\mu_2$, which are the rest masses of the shells.
\begin{eqnarray} \label{mu1}
\mu_1&=&-R_1\left[ (1-\frac{2M}{R_1}+\dot{R_1}^2)^{\frac{1}{2}}-(1-\frac{2m}{R_1}+\dot{R_1}^2)^{\frac{1}{2}}\right] ,\\
\label{mu2}
\mu_2&=&-R_2\left[ (1-\frac{2m}{R_2}+\dot{R_2}^2)^{\frac{1}{2}}-(1+\dot{R_2}^2)^{\frac{1}{2}}\right] .
\end{eqnarray}
 Here, the dot represents the derivative with respect to the proper time of an observer who is sitting on the shell. The shells are assumed to have no pressure. From Eq.~(\ref{mu1}) and (\ref{mu2}), we have
\begin{eqnarray} \label{trajectory1}
\dot{R_1}&=&\Big( \frac{(M-m)^2}{\mu_1^2}-1+\frac{M+m}{R_1}+\frac{\mu_1^2}{4R_1^2}\Big)^{\frac{1}{2}}\\
\label{trajectory2}
\dot{R_2}&=&\Big( \frac{m^2}{\mu_2^2}-1+\frac{m}{R_2}+\frac{\mu_2^2}{4R_2^2}\Big)^{\frac{1}{2}}
\end{eqnarray}
Then, the proper times on the shells are given by
\begin{eqnarray}
\tau_1 &=& \int \frac{dR_1}{\dot{R_1}}\\
\tau_2 &=& \int \frac{dR_2}{\dot{R_2}}
\end{eqnarray}
The coordinate time of an observer on $S_1$ is
\begin{eqnarray}
\label{out}
t_I (\tau_1,R_1)&=&\int \frac{\Big(1+\frac{\dot{R}_1^2}{1-\frac{2M}{R_1}}\Big)^\frac{1}{2}}{\Big(1-\frac{2M}{R_1}\Big)^\frac{1}{2}}d\tau_1\\
\label{out1}
t_{II}(\tau_1,R_1)&=&\int \frac{\Big(1+\frac{\dot{R}_1^2}{1-\frac{2m}{R_1}}\Big)^\frac{1}{2}}{\Big(1-\frac{2m}{R_1}\Big)^\frac{1}{2}}d\tau_1
\end{eqnarray}
The coordinate time of an observer on $S_2$ is
\begin{eqnarray}
\label{in}
t_{II} (\tau_2,R_2)&=&\int \frac{\Big(1+\frac{\dot{R}_2^2}{1-\frac{2m}{R_2}}\Big)^\frac{1}{2}}{\Big(1-\frac{2m}{R_2}\Big)^\frac{1}{2}}d\tau_2\\
\label{in1}
t_{III}(\tau_2,R_2)&=&\int \Big(1+\dot{R}_2^2\Big)^\frac{1}{2}d\tau_2
\end{eqnarray}
This fixes the geometry of the problem. We can easily reconstruct the single shell case by setting $m=\mu_2=0$.

In the following sections, our goal will be to calculate the momentum energy tensor for radiation from a collapsing object as defined in \cite{Davies}  and described in detail in the book \cite{1984qfcs.book.....B}. The original space-time is  $(3+1)$- dimensional, but if one considers a spherically symmetric case, it effectively reduces to a $(1+1)$- dimensional space-time $(t,r)$.  This reduced space-time still contains information about the non-trivial curvature around the collapsing object and faithfully reproduces Hawking radiation. It is true that one loses the information about the angular momentum barrier, but we know that this barrier just modifies the original Hawking radiation by generating the so-called greybody factors, which will not be relevant for our discussion.

\section{Scalar field propagating in the background of the collapsing shells}

In this section we add a massless scalar field which propagates in the background of this collapsing system. For simplicity, we assume that the shells are transparent to the scalar field, which is a usual assumption. The evolution of the scalar field in this curved background is described by
\begin{equation}
\Box \psi =0
\end{equation}
where the $\Box$ operator is covariant. The $\Box$ operator must be calculated separately in three different regions I, II and III.

\begin{eqnarray}
&& \partial_{t_I}^2 \psi- \frac{1}{r^2}\partial_{r_1^*}(r^2\partial_{r_1^*}\psi ) +V(r,M)\psi=0 \mbox{, for $r>R_1$}\\
&& \partial_{t_{II}}^2 \psi- \frac{1}{r^2}\partial_{r_2^*}(r^2\partial_{r_2^*}\psi ) +V(r,m)\psi=0 \mbox{, for $R_1>r>R_2$} \nonumber \\
&& \partial_{t_{III}}^2 \psi- \frac{1}{r^2}\partial_{r}(r^2\partial_{r}\psi ) +V(r,0)\psi=0 \mbox{, for $R_2>r$}\nonumber \\
&&V(r,M)=\frac{L^2}{(1-\frac{2M}{r})r^2}\nonumber
\end{eqnarray}
In the region III, the space-time is Minkowski by the Birkhoff's theorem.
The starred radial coordinates $r_1^*=\int \frac{dr}{1-\frac{2M}{r}}$ and $r_2^*=\int \frac{dr}{1-\frac{2m}{r}}$ are the usual tortoise coordinates. $L^2$ is the eigenvalue of the angular part of the equation. If one performs a change of variables $\psi=\phi /r$ (angular part is ignored here), the above equations reduce to

\begin{eqnarray}
&& \partial_{t_I}^2 \phi- \partial_{r_1^*}^2\phi  +F(r_1^*,r,M)\phi=0 \mbox{, for $r>R_1$}\\
&& \partial_{t_{II}}^2 \phi- \partial_{r_2^*}^2 \phi +F(r_2^*,r,M)\phi=0 \mbox{, for $R_1>r>R_2$}\nonumber \\
&& \partial_{t_{III}}^2 \phi- \partial_r^2 \phi+F(r,r,0)\phi =0 \mbox{, for $R_2>r$} \nonumber \\
&&F(r^*,r,M)=r\partial_{r^*}^2r+rV(r,M)  \nonumber
\end{eqnarray}

 The function $F$ in the above equations is an effective potential barrier. Since there are only two independent variables in each of these equations (the time and radial coordinate), they effectively represent a 2-dimensional penetration problem (similar analysis can be found in \cite{1984qfcs.book.....B} section 8.1 and 8.2). If we are interested in the solution far away from the potential barrier, $F$, the solution reduces to a regular $1+1$ dimensional plane wave. The effect of the potential barrier is generally treated as the graybody factor, and it will not be important for our analysis here. Therefore our problem reduces to a $1+1$-dimensional free scalar field which satisfies the wave equation

\begin{eqnarray}
&& \partial_{t_I}^2 \phi- \partial_{r_1^*}^2\phi =0 \mbox{, for $r>R_1$}\\
&& \partial_{t_{II}}^2 \phi- \partial_{r_2^*}^2 \phi =0 \mbox{, for $R_1>r>R_2$}\nonumber \\
&& \partial_{t_{III}}^2 \phi- \partial_r^2 \phi=0 \mbox{, for $R_2>r$}\nonumber
\end{eqnarray}

 The trajectory of the spherical shell is given by Eq.~(\ref{trajectory1}) and (\ref{trajectory2}). There are two solutions to the wave equation for $r> R_1$. One is the ingoing mode
\begin{equation}
\phi_{\rm in} \sim \exp(-i\omega v)
\end{equation}
and the other one is the outgoing mode
\begin{equation}
\phi_{\rm out} \sim \exp(-i\omega p(u)) ,
\end{equation}
where we defined the ingoing and outgoing null coordinates $v=t+r^*$ and $u=t-r^*$.

\begin{figure}
   \centering
\includegraphics[width=5cm]{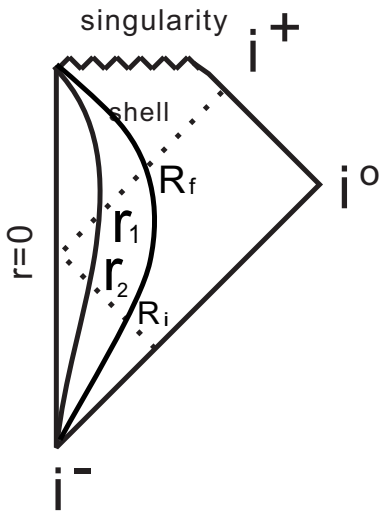}
\caption{Penrose diagram for the transparent collapsing shells. The incoming scalar mode crosses the two shells at locations $R_i$, $r_2$ on its way in, reaches the center, becomes the outgoing mode, and crosses the two shells again at locations $r_1$ and $R_f$ respectively.}
\label{Ref}
\end{figure}

The shells in our discussion here are massive. While the shells are collapsing, the incoming scalar field mode passes through $S_1$ and $S_2$ at locations $r=R_i$ and $r=r_2$ respectively. Once it passes through the center, it becomes an outgoing mode. It passes again through $S_2$ and $S_1$ at locations $r=r_1$ and $r=R_f$ respectively, as shown in Fig.~\ref{Ref}. Since the massless scalar field moves at the speed of light, it must satisfy the condition
\begin{eqnarray}
\label{con1}
&&t_{II}(R_f)-t_{II}(r_1)= r_2^*(R_f)-r^*_2(r_1) \\
\label{con2}
&&t_{III}(r_1)-t_{III}(r_2)= r_1+r_2 \\
\label{con3}
&&t_{II}(r_2)-t_{II}(R_i)= r^*_2(R_i)-r_2^*(r_2)
\end{eqnarray}

\begin{figure}
   \centering
\includegraphics[width=9cm]{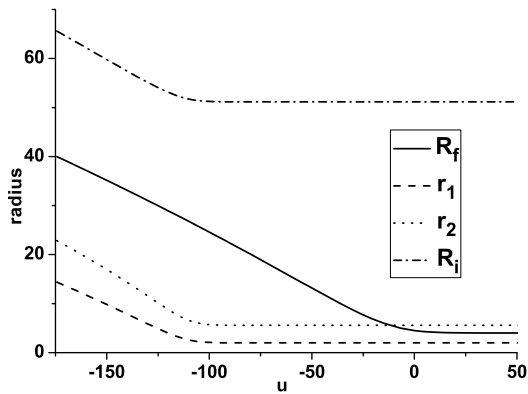}
\caption{Radii of the shells $R_i$, $r_2$, $r_1$ and $R_f$  at the moments when the scalar field mode crosses them on its way in and out. To get concrete numbers we set $u=t(R_f)-r_1^*(R_f )$, $m=\mu_1=\mu_2=M/2 =1$ and $\Delta =100$ (the parameter $\Delta$ is defined in Eq.~(\ref{delta})). }
\label{radius}
\end{figure}

The shells' radii at the moments of these four crossings are shown in Fig.~\ref{radius}.

When the scalar wave mode comes out of the outer shell we have $u=t(R_f)-r_1^*(R_f)$, but $p=v(R_i)=t(R_i)+r_1^*(R_i)$. The function $p$ can be written in terms of the variable $u$ with the help of Eq.~(\ref{con1}), (\ref{con2}) and  (\ref{con3}).

In the concrete numerical computations, we first give a particular value for $R_f$. With this value, $r_1$ is obtained from Eq.~\ref{con1}. Then $r_2$ is obtained by substituting the value for $r_1$ into Eq.~\ref{con2}. Similarly, $R_i$ could be obtained by substituting $r_1$ into Eq.~\ref{con3}. The relation between $R_i$ and $R_f$ can be written as $R_i(r_2(r_1(R_f)))$. Now $v(R_i)$ can be written as a function of the variable $R_f$ (or equivalently $u(R_f)$). When we compute $R_i$ as a function of $R_f$, we set the digital precision $40 $ to avoid numerical errors near the horizon.

\section{Comparison of energy fluxes}

In this section we calculate the energy flux coming from the two collapsing shells considered in the previous section, and compare it to the flux emitted by a single shell of the equivalent mass.
The energy flux for a collapsing object of any given gravitational trajectory is defined in \cite{Davies,1984qfcs.book.....B}. In a $1+1$ dimensional $(u,v)$ spacetime, it can be written as
\begin{eqnarray}
T_{uu}&=&<T_{uu}^B>+\frac{1}{24\pi}\Bigg(\frac{3}{2}\Big(\frac{p''}{p'}\Big)^2-\frac{p'''}{p'}\Bigg)\\
T_{uv}&=&<T_{uv}^B>\\
T_{vv}&=&<T_{vv}^B>  .
\end{eqnarray}
Primes indicates derivative with respect to the coordinate $u$. The components  $<T_{\alpha \beta}^B>$ refer to the energy momentum density in the Boulware vacuum.  These terms will not travel to infinity and therefore are irrelevant for us (we are concerned about what a distant outside observer would see). Only the second term in $T_{uu}$ will survive at $r\rightarrow \infty$. To get concrete numerical results, for simplicity we set $\mu_1=\mu_2 =m=M/2 =1$. We also set up the condition so that the two shells do not cross each other during the collapse. We can find the coordinates $t_I$, $t_{II}$ and $t_{III}$ from Eqs.(\ref{out}), (\ref{out1}), (\ref{in}) and (\ref{in1}). There will be an arbitrary integration constant in these integrals. Suppose that $S_2$ arrives at location $R_2=R$ at the moment $t_{II}(\tau_2,R)$, while  $S_1$ arrives at location $R_1=R$ at the moment $t_{II}(\tau_1,R)$. The time interval
 \be \label{delta}
 \Delta \equiv t_{II}(\tau_1,R)-t_{II}(\tau_2,R)
 \ee
  is the time difference between the moments when these two shells cross through $r=R$. We set this to be at $R=5$, which removes one of the integration constants. This is equivalent to choosing the initial conditions for the equations of motion.

\begin{figure}
   \centering
\includegraphics[width=9cm]{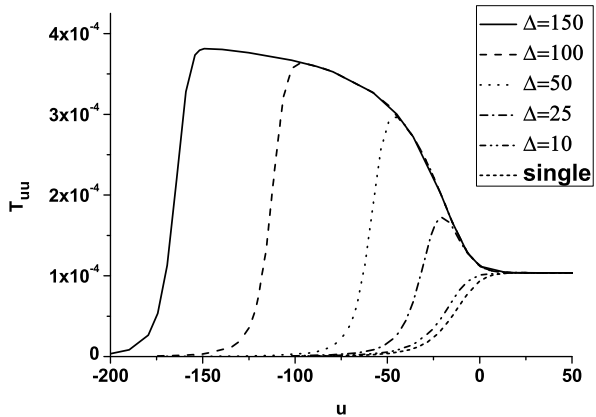}
\caption{The curves represent the energy flux $T_{uu}$ at $r\rightarrow \infty$ as a function of time coordinate $u$, for several values of the time interval $\Delta$ between the moments when these two shells cross through $r=5$. $\mu_1=\mu_2 =m=M/2 =1$. For comparison, we plot the case of the single shell with mass $M=\mu_1=2$ and $m=\mu_2=0$ (the last curve on the right).  At late enough time the fluxes become indistinguishable, and equal the flux emitted from a static black hole $\frac{1}{3072\pi}$. We clearly see that the flux emitted by a single shell of the rest mass $\mu_1=2$ is different from the total flux emitted by two shells of the rest mass $\mu_1=\mu_2=1$. }
\label{flux}
\end{figure}

The main result of our analysis is shown in Fig.~\ref{flux}. We plotted the term $T_{uu}$, which is just the energy flux emitted by the system of two shells as seen at infinity, as a function of time coordinate $u$. We set $G=c=\hbar=k=1$.
The system of two shells has more freedom than one shell, which we parameterize with the time interval $\Delta$ which represents the difference between the moments when these two shells cross through some fixed radius, which we chose to be $r=5$. For comparison, we add the plot of the single shell with mass equal to the mass of the two shells.
We clearly see that the flux emitted by a single shell of the rest mass $\mu_1=2$ is different from the total flux emitted by two shells of the rest mass $\mu_1=\mu_2=1$.
Thus, the flux of pre-Hawking radiation at infinity can in principle reveal the mass distribution of the collapsing system.

One may also notice that the flux emitted by two shells is higher than that from the static black hole (plateau on the right in Fig.~\ref{flux}). The reason is that the inner shell $S_2$ reaches its own Schwarzschild radius first and generates flux ($\frac{1}{768\pi}$) with higher temperature. After the flux peaks, it gradually decreases because the temperature decreases when $S_1$ reaches the Schwarzschild radius of the system.

\section{Conclusion}
Physical properties of a black hole are completely determined by its mass, charge and angular momentum. There is no additional information left after a black hole is formed. Black holes made of matter with different mass distribution will emit exactly the same Hawking radiation. Then one cannot recover any additional information with conventional physics if only Hawking radiation from a static black hole is considered. Either quantum effects are able to remove the singularity (and global event horizon), or some strongly non-local physics transfers the information from inside to outside the horizon \cite{Saini:2014qpa,Greenwood:2008ht,Bogojevic:1998ma,Lowe:2006xm,Lowe:1999pk,Giddings:2006sj}. Both of these options would significantly diverge from the conventional black hole physics. Therefore, it is still useful to explore what happens if we re-formulate the problem a bit.

The most conservative solution to this problem would imply that information is released during the collapse before the static horizon is formed. Whether this information is extracted before or after the horizon is formed, ultimately does not matter, since the outside observer observes only the asymptotic ``in" and ``out" states (i.e. the collapsing object and the outgoing radiation). It is well known that a collapsing object can shed its higher multipole moments in the form of gravity waves during the so-called balding phase before reaching a perfect spherically symmetric form. However, a non-trivial mass distribution which is still spherically symmetric will not excite gravity waves (at least not classically). Thus, it is very important to extend this balding phase to this case.

In this paper, we considered a specific problem of possible reconstruction of the mass distribution from radiation which is emitted during the collapse. We considered
two different initial states which have the same mass. One state is a single collapsing spherically symmetric shell of a certain mass, while the second state may be made of two concentric shells whose total  mass is equal to the mass of the single shell in the first case. The final states in both cases are the same - a static black hole with the same total mass. We concentrated on the pre-Hawking radiation which is emitted before the static black hole is formed, but we used the standard analysis of tracing the field modes in the time-dependent gravitational background. We calculated the gravitational trajectory for these two cases, and then the components of energy momentum tensor in
a $(1+1)$-dim spherically symmetric space-time defined by $(t,r)$. We showed that the flux of energy emitted by a single shell is notably different from the flux emitted by two concentric shells of the equivalent total mass. This implies that an observer studying the flux of particles from a collapsing object could in principle reconstruct information  about the mass distribution of the collapsing object.

\begin{acknowledgments}
D.C. Dai was supported by the National Science Foundation of China (Grant No. 11433001 and 11447601), National Basic Research Program of China (973 Program 2015CB857001), No.14ZR1423200 from the Office of Science and Technology in Shanghai Municipal Government, the key laboratory grant from the Office of Science and Technology in Shanghai Municipal Government (No. 11DZ2260700) and  the Program of Shanghai Academic/Technology Research Leader under Grant No. 16XD1401600. DS partially supported by the US National Science Foundation, under Grant No. PHY-1417317.

\end{acknowledgments}

\end{document}